\documentclass[iop]{emulateapj}
\usepackage{amssymb}
\usepackage{graphicx}
\usepackage{epsfig}
\usepackage{amsmath}
\usepackage[colorlinks=true,  citecolor=blue,  
  linkcolor=blue,  menucolor=blue, urlcolor=blue,
  linkbordercolor={0 0 1}, frenchlinks=True, breaklinks]{hyperref}
\usepackage{natbib}
\usepackage{booktabs}
\usepackage{bookmark}
\usepackage{enumerate}
\usepackage{multirow}
\newcommand{\myemail}{ylgao@mail.ustc.edu.cn, ecwang16@ustc.edu.cn, xkong@ustc.edu.cn}

\newcommand{\ha}{\hbox{H$\alpha$}}
\newcommand{\hb}{\hbox{H$\beta$}}

\newcommand{\oii}{\hbox{[O\,{\sc ii}]}}
\newcommand{\nii}{\hbox{[N\,{\sc ii}]}}
\newcommand{\sii}{\hbox{[S\,{\sc ii}]}}
\newcommand{\oiii}{\hbox{[O\,{\sc iii}]}}

\newcommand{\hii}{\hbox{H\,{\sc ii}}}

\shorttitle{What determines the local metallicity?}
\shortauthors{Y.L. Gao et al.}

\begin{document}

\title{What determines the local metallicity of galaxies: global stellar mass, local stellar mass surface density or star formation rate?}

\author{Yulong Gao\altaffilmark{1,2}, Enci Wang\altaffilmark{1,2,3}, Xu Kong\altaffilmark{1,2}, Zesen Lin\altaffilmark{1,2}, Guilin Liu\altaffilmark{1,2}, Haiyang Liu\altaffilmark{1,2}, Qing Liu\altaffilmark{1,2}, Ning Hu\altaffilmark{1,2},  Berzaf Berhane Teklu\altaffilmark{1,2}, Xinkai Chen\altaffilmark{1,2}, and Qinyuan Zhao\altaffilmark{1,2}}
\affil{$^{1}$ CAS Key Laboratory for Research in Galaxies and Cosmology, Department of Astronomy, University of Science and Technology of China, Hefei 230026, China; \myemail \\
$^{2}$ School of Astronomy and Space Sciences, University of Science and Technology of China, Hefei 230026, China \\
$^{3}$ Department of Physics, Institute for Astronomy, ETH Zurich, CH-8093 Zurich, Switzerland}

\begin{abstract}
    The metallicity and its relationship with other galactic properties is a fundamental probe of the evolution of galaxies. In this work, we select about 750,000 star-forming spatial pixels from 1122 blue galaxies in the MaNGA survey to investigate the global stellar mass -- local stellar mass surface density -- gas-phase metallicity ($M_*$ -- $\Sigma_*$ -- $Z$ ) relation. At a fixed $M_*$, the metallicity increases steeply with increasing $\Sigma_*$. Similarly, at a fixed $\Sigma_*$, the metallicity increases strongly with increasing $M_*$ at low mass end, while this trend becomes less obvious at high mass end. We find the metallicity to be more strongly correlated to $\Sigma_*$  than to $M_*$. Furthermore, we construct a tight (0.07 dex scatter) $M_*$ -- $\Sigma_*$ -- $Z$ relation, which reduces the scatter in the $\Sigma_*$ -- $Z$ relation by about 30$\%$ for galaxies with $7.8 < {\rm log}(M_*/M_\odot) < 11.0$, while the reduction of scatter is much weaker for high-mass galaxies. This result suggests that, especially for low-mass galaxies, the $M_*$ -- $\Sigma_*$ -- $Z$ relation is largely more fundamental than the $M_*$ -- $Z$ and $\Sigma_*$ -- $Z$ relations, meaning that both $M_*$ and $\Sigma_*$ play important roles in shaping the local metallicity.  We also find that the local metallicity is probably independent on the local star formation rate surface density at a fixed $M_*$ and $\Sigma_*$. Our results are consistent with the scenario that the local metallicities in galaxies are shaped by the combination of the local stars formed in the history and the metal loss caused by galactic winds.
\end{abstract}

\keywords{galaxies: abundances -— galaxies: evolution -— galaxies: starburst -— star formation}

\section{Introduction}
\label{sec:intro}
The chemical processes of the gas and stars in galactic environments is of key importance in understanding the formation and evolution of galaxies, and the metallicity of the interstellar medium ($Z$), along with its relationship with other galactic properties, is a fundamental probe. The observed metallicity is a result of the interplay between the enrichment from previous generations of stars, the metal loss caused by winds from stars and active galactic nuclei (AGN), and the dilution by metal-poor gas inflows. In spite of the investigation for the chemical evolution is complex, the scaling relations between metallicity and other galactic properties provides a powerful alternative approach. 

Since its establishment by \cite{Lequeux1979a}, the relation between stellar mass and metallicity has been studied for decades \citep[e.g.][]{Garnett1987a,Tremonti2004}. It is a positive correlation, meaning that the metallicities of galaxies increase with increasing stellar masses. The significant scatter in the stellar mass -- metallicity relation (MZR) implies the complication of the underlying physical and chemical processes in an evolving galaxy. Variations in  galactic parameters, such as the star formation rate (SFR), the 4000 $\mathrm{\AA}$ break ($D_{\rm n} 4000$), and the galaxy size and morphology, are found to contribute to the scatter in MZR \citep[e.g.][]{Ellison2008,Mannucci2010a,Yates2012,Andrews2013,Lian2015,Bothwell2016,Wang2018}. 
In particular, using the SDSS Data Release (DR) 7 \citep{Abazajian2009}, \cite{Mannucci2010a} and \cite{Andrews2013}  found that metallicity to be anti-correlated with SFR at a fixed stellar mass, and the scatter in MZR is siginificantly suppressed when a stellar mass -- SFR -- metallicity relation (fundamental metallicity relation; FMR) is constructed. Nevertheless, with the same data SDSS DR7, \cite{Salim2014} argued that a more physically motivated second parameter for the MZR is the specific SFR (sSFR) when ${\rm log}(M_*/M_\odot) \leq 10.5$.  \cite{Bothwell2016} derived the molecular hydrogen masses from the CO luminosity for 221 galaxies at 0 $< z <$ 2, then suggested that instead of the SFR, the third parameter that should be introduced into FMR is the gas mass.  

The MZR and FMR are both established primarily using global galaxy parameters, it is also important to understand whether there is a more fundamental relation to probe the global MZR or FMR with local galactic parameters, e.g. stellar mass surface density, local metallicity, local SFR and local $D_{\rm n}4000$.  The local metallicity vs. local stellar mass surface density relation has been investigated for years. \citet{Edmunds1984a} and \citet{Vila-Costas1992a} reported that $\hii$ regions with larger stellar mass densities are more metal-rich than those with lower densities. \cite{Moran2012} performed long-slit spectroscopy on 174 star-forming galaxies and found a correlation between the local stellar mass surface density and metallicity, an analog to the global MZR. More recently, thanks to the emergence of the integral field spectroscopy (IFS) technique, analyzing these relations in a spatially resolved manner in relatively large samples of galaxies becomes feasible. \citet{Rosales-Ortega2012} demonstrated the existence of a local relation between stellar mass surface density, metallicity, and SFR density using 38 nearby galaxies from the PINGS survey \citep{Rosales-Ortega2010} and the CALIFA survey \citep{Sanchez2012}.  Based on the SDSS-IV MaNGA survey \citep{Bundy2015,Yan2015,Yan2016,Blanton2017}, \cite{Barrera-Ballesteros2016a} and \cite{Zhu2017} presented a tight local stellar mass surface density versus metallicity relation, suggestive of the fact that local properties play a key role in determining the metallicity in typical disk galaxies. These authors also reproduced the global MZR using the local relation, concluding that the global relation is a scale-up, integrated effect of the local relation. In addition, a series of studies with IFS data explored whether there is a secondary dependent parameter in the global MZR or the stellar mass surface density -- metallicity relation. Using the data in CALIFA survey, \cite{Sanchez2013a} and \cite{Sanchez2017a} presented the local mass surface density -- metallicity relation and global MZR, and confirmed the nonexistence of such these relations with (s)SFR. \cite{Barrera-Ballesteros2017} also demonstrated that the MZR is independent on the (s)SFR with more than 1700 galaxies in MaNGA survey. 

It is well known that the global stellar mass dominates the metallicity in the center of galaxies \citep{Tremonti2004,Mannucci2010a}. However, it remains unclear whether the global stellar mass has an nonnegligible effect on shaping the distribution of local metallicity. Intuitively, higher global stellar mass helps producing deeper gravitational potential wells in a massive galaxy that retain the metals (re)processed by stars, possibly leading to a higher local metallicity than in a less massive galaxy, even if these galaxies have an identical surface density distribution of stellar mass. \cite{Barrera-Ballesteros2016a} have found that the stellar mass surface density and metallicity relation is largely independent on the total stellar mass for massive (${\rm log}(M_*/M_\odot) \geq 9.5$) galaxies.  In this work, we will investigate whether the global stellar mass, especially for low-mass galaxies, is a potential parameter that participates in controlling local metallicity, and to construct a correlation between global stellar mass, local stellar mass surface density and local metallicity, in order to shed light on understanding the local chemical evolution history in galaxies. 

The remainder of this paper is organized as follows. In Section \ref{sec:data}, we present the sample selection for star-forming galaxies, the methodology for detecting and measuring nebular emission lines, and the determination of dust attenuation properties, gas-phase oxygen abundances and other galaxy parameters. In Section \ref{sec:results}, we describe the methods for estimating the strength of the correlation between $M_*$ -- $Z$  and $\Sigma_*$ -- $Z$, as well as that of a new $M_*$ -- $\Sigma_*$ -- $Z$ relation. We discuss our results in the context of the literature, and the dependence of the residuals on other parameters in Section \ref{sec:discussion}. Finally, we summarize our results in Section \ref{sec:summary}. Throughout this paper, we adopt a flat $\Lambda$CDM cosmology with $\Omega_\Lambda=0.7$, $\Omega_{\rm m}=0.3$, and $H_0=70$ km s$^{-1}$ Mpc$^{-1}$.
 
\section{Data Analysis}
\label{sec:data}

    \subsection{MaNGA Overview}
    \label{subsec:manga}

    The MaNGA (Mapping Nearby Galaxies at Apache Point Observatory) survey, one of the three core programs in the Sloan Digital Sky Survey IV \citep[SDSS--IV,][]{Blanton2017}, is an IFS survey aiming at 10,000 nearby galaxies that are selected from the NASA-Sloan-Atlas (NSA) catalog \footnote{http://www.nsatlas.org} \citep{Blanton2011,Bundy2015}. The redshift of these target galaxies span a range of $0.01<z<0.15$. The spectrographs of the MaNGA survey provide a spectral coverage of 3600 --  10300 $\mathrm{\AA}$ at a resolution of $R \sim $ 2000 \citep{Drory2015}. The diameter of an individual fiber is 2\arcsec. SDSS DR14 \citep{Abolfathi2018}, the second data release of MaNGA, has delivered a public sample of 2812 galaxies with spatially resolved IFS mapping \footnote{http://www.sdss.org/dr14/manga/manga-data/catalogs}. In this work, we treat these 2812 galaxies as the parent sample.  

    \subsection{Spectral Fitting And Emission-line Measurements}
    \label{subsec:spec-fitting}

    In general, one needs to disintegrate the emission lines from the underlying stellar continuum, and then measure the fluxes of these lines \citep{Hu2016,Ylgao2017}. For strong emission lines, the subtraction of the stellar continuum has negligible effect on the flux measurements. In this work, this decomposition is performed using the \textsc{STARLIGHT} code \citep{CidFernandes2005}. We fit each spectrum with the combination of 45 single stellar populations (SSPs) from \citet{Bruzual2003} model, assume a \citet{Chabrier2003} initial mass function (IMF), and obtain the stellar population parameters. These SSPs are evenly distributed on an age -- metallicity grid, which consists of 15 ages (ranging from 1 Myr to 13 Gyr) and 3 different metallicities (Z = 0.01, 0.02, 0.05). We also mask out optical emission lines using the standard emission line masks of \textsc{STARLIGHT}. 
    
    Before the spectral fitting, these spectra all have been corrected for Galactic extinction using the color excess $\rm E(B-V)$ map of the Milky Way \citep{Schlegel1998}, and the \citet{Calzetti2000} attenuation law is adopted for correcting their intrinsic stellar reddening.  We calculate the fluxes of the strong emission lines (e.g. $\oii\lambda3727$, $\oiii\lambda4363$,  $\hb$, $\oiii\lambda\lambda4959,5007$, $\ha$, $\nii\lambda\lambda6548,6583$, $\sii\lambda\lambda6717,6731$) by fitting their profiles to multiple Gaussians using the IDL package \textsc{mpfit} \citep{Markwardt2009}. The signal-to-noise ratios (S/N) of these emission lines are estimated following \citet{Ly2014}. 

    Comparing the Balmer decrement (e.g. $\ha/\hb$ ratio) to their intrinsic values that can be theoretically calculated under the ``Case B'' assumption, one could derive the dust attenuation (the dust geometry is postulated to be a non-scattering screen at a distance from the emitter, see \citet{Liu13} for discussion on the complication of the dust geometry in galaxies). The Balmer decrement is insensitive to the temperature and electron density, and here we adopt a intrinsic flux ratio of $(\ha/\hb)_0=2.86$ \citep{Hummer1987}. We use the \cite{Calzetti2000} reddening formalism to derive the color excesses $\rm E(B-V)$, and correct for the dust extinction for the emission line fluxes. When the observed flux ratio is unphysically $\ha/\hb < 2.86$ (due to error scatter), we set $\rm E(B-V)$ to zero in an ad hoc manner.

    \begin{figure*}
        \begin{center}
        \includegraphics[width=0.9\textwidth]{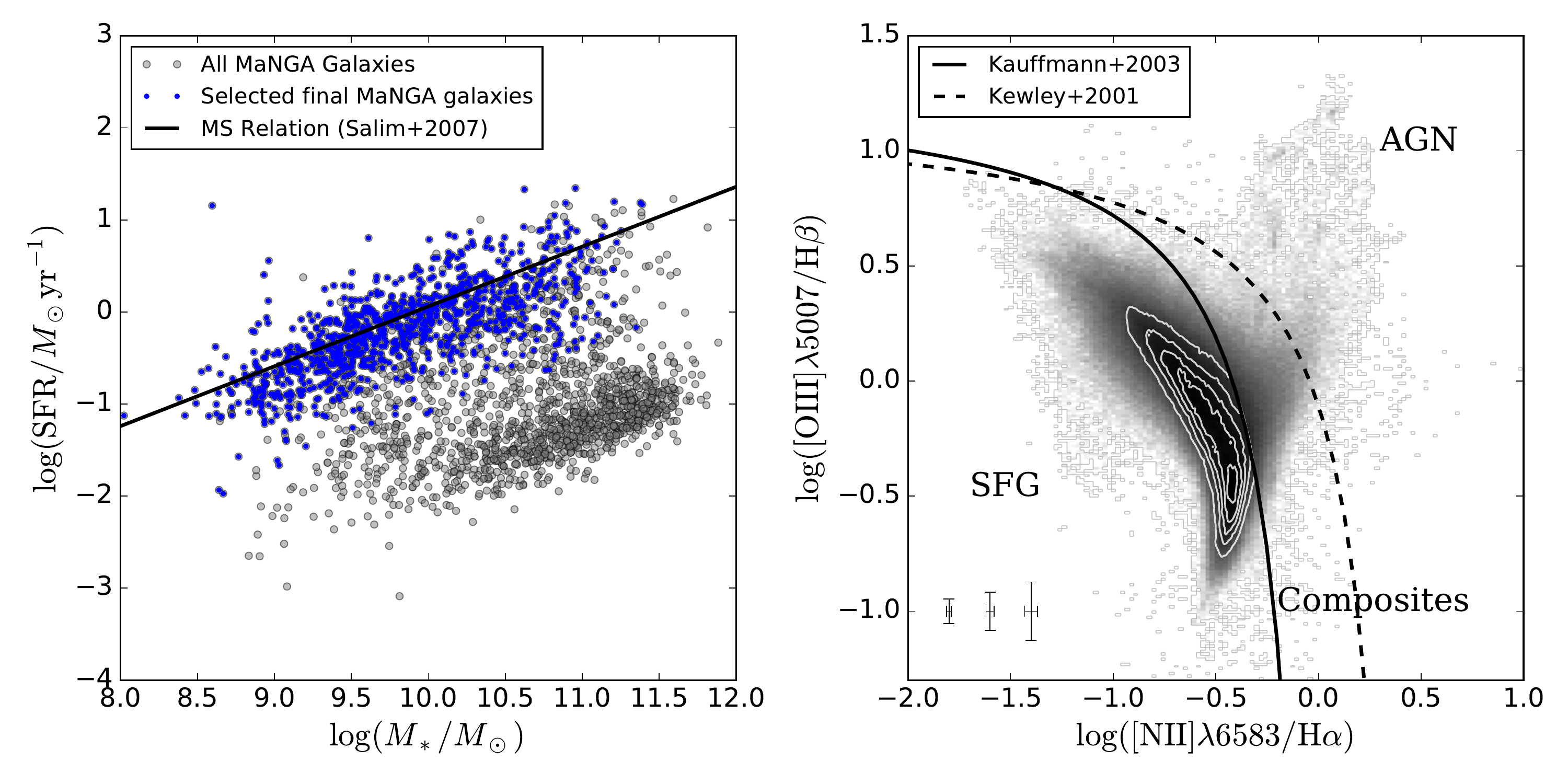}
        \end{center}
        \caption{$Left$: The total stellar mass -- SFR (main sequence) relation for MaNGA galaxies. The blue points represent our final 1122 sample galaxies, while the grey points show all of the galaxies in MaNGA data. The main sequence relation for star-forming galaxies in local universe \citep{Salim2007,Speagle2014} is shown as the black solid line.   $Right$: The BPT diagram for all spaxels in our sample blue galaxies. The grayscale 2D histogram shows the number density of observations. The white contours cover the $85\%$ of our final sample spaxels. The solid and dashed lines are the demarcation curves between SFGs and AGNs defined by \cite{Kauffmann2003} and \cite{Kewley2001}, respectively. The errorbars in the left bottom corner represent the error values at the distribution of $16\%$, $50\%$ and $84\%$, respectively. }
        \label{fig:bpt_manga}
    \end{figure*}

    \subsection{Sample Selection}
    \label{subsec:sample}
    
    We retrieve the photometry of these 2812 galaxies from the NSA catalog. Since the metallicity under consideration is gas-phase metallicity, we focus on the star-forming galaxies with ${\rm NUV} - r < 4$ \citep{Li2015}, whose metallicities can be reliably measured. Due to the same reason, we only consider the $\hii$ regions of the selected galaxies, confining the spatial pixels (a.k.a. ``spaxels'') to those with S/N($\ha$) $>$ 5, S/N($\hb$) $>$ 5, S/N($\oiii\lambda3727$) $>$ 5, S/N($\oiii\lambda\lambda4959,5007$) $>$ 5, S/N($\nii \lambda 6583$) $>$ 3 and equivalent width (EW) of $\ha$ larger than 10 $\rm \AA$. Meanwhile, in order to obtain reliable stellar mass surface density, we select the spectra with a continuum S/N higher than 3 at 5500 $\mathrm{\AA}$. We exclude the spaxels affected by the active galactic nucleus (AGN) using the \cite{Kauffmann2003} demarcation line in the BPT diagram \citep{Baldwin1981,Kewley2001,Kauffmann2003}. We also exclude these galaxies, in which the selected spaxels number is less than 10$\%$ of the spaxels in this galaxy.
    
    As a result, we achieve 1122 blue galaxies with about 750,000 useful spaxels. In Figure \ref{fig:bpt_manga}, the $left$ panel shows total stellar mass -- SFR (main sequence) relation for MaNGA galaxies, in which the $M_*$ and SFR are obtained from MPA/JHU catalogue \footnote{http://www.sdss.org/dr12/spectro/galaxy$\_$mpajhu/} \citep{Kauffmann2003,Salim2007}. The blue points represent our final 1122 sample galaxies, while the grey points show all of the galaxies in MaNGA data. The main sequence relation for star-forming galaxies in local universe \citep{Salim2007,Speagle2014} is shown as the black solid line. The $right$ panel shows the BPT diagram constructed with all spaxels in our sample blue galaxies, in which the grayscale two-dimensional histogram demonstrates the number density (crowdedness) of spaxels and the white contours cover the $85\%$ of our final sample spaxels.  The solid and dashed lines are the demarcations between star-forming galaxies (SFGs) and AGNs defined by \cite{Kauffmann2003} and \cite{Kewley2001}, respectively. 

    \subsection{Determinations of Metallicity and Other Physical Properties}
    \label{subsec:metallicity}

    The electron temperature ($T_e$) method is generally deemed as the most reliable approach for deriving metallicity, which is based on the ratios of faint auroral-to-nebular emission lines \citep{Lin2017}, such as $\oiii\lambda4363/\oiii\lambda5007$. However, only a handful of spaxels show pronounced $\oiii\lambda4363$ lines. Besides, variety of calibrators can also be used to estimate the metallicity \citep{Kewley2008}. Based on the photoionization models for $\hii$ regions, these emission line ratios, like $(\oii\lambda3727+\oiii\lambda\lambda4959,5007)/\hb$ \citep[R23,][]{Kobulnicky04}, $\nii\lambda6583/\oii\lambda3727$ \citep[N2O2,][]{Kewley2002}, can be used as the calibrators to reproduce the metallicity.  Some other diagnostics, like $(\oiii\lambda5007/\hb)/(\nii\lambda6583/\ha)$ and $\nii\lambda6583/\ha$ \citep[O3N2, N2,][]{Pettini2004}, are calibrated by empirical fitting to the electronic temperature ($T_e$) method with strong-line ratios for $\hii$ regions and galaxies. However, there is a large discrepancy between the metallicities derived by different calibrators \citep{Kehrig2013,Morisset2016}. For example, compared with the N2O2 diagnostic, O3N2 and N2 almost couldn't reproduce the super-solar oxygen abundances and will derive lower metallicities \citep{Blanc2015,Zhang2017}.

    In this work, we adopt the N2O2 diagnostic to determine the oxygen abundances. The N2O2 index introduced in \cite{Dopita2000} and improved by \cite{Dopita2013}, is defined as
    \begin{equation}
    \rm N2O2 \equiv log(\frac{\nii\lambda6583}{\oii\lambda\lambda3727,3729}),
    \end{equation}
    and is converted to metallicity using the relation in \cite{Kewley2002}  as
    \begin{equation}
    \begin{split}
    \rm 12 + log(O/H) = log(1.54020 + 1.26602 \times N2O2 \\
     \rm + 0.167977 \times N2O2^2)+ 8.93,
    \end{split}
    \end{equation}
    with a typical error of 0.04 dex when $\rm 12 + log(O/H) \geq 8.6$. In order to obtain the general results that are independent on the metallicity calibrators, we also calculate the metallicities with O3N2 and N2 indices in Appendix.   

    The global stellar mass $M_*$ is retrieved from the NSA catalog \citep{Blanton2011}. In order to derive the local stellar mass surface density $\Sigma_*$, we divide the local stellar mass in each spaxel of the output from the \textsc{STARLIGHT} fits by its corresponding physical area, following the method in \cite{Barrera-Ballesteros2016a} to correct for the inclination by applying the minor-to-major axis ratio ($b/a$) retrieved from the NSA catalog. The uncertainty of the local stellar mass for each spaxel is smaller than 0.11 dex when the S/N of spectrum continuum is larger than 5. Comparing our mass surface density $\Sigma_*$ with the public MaNGA Pipe3D value added catalog \citep{Sanchez2016a}, we note that our $\Sigma_*$ is systematically lower than the Pipe3D values, with a offset of 0.539 dex with a scatter of 0.22 dex. The offset is likely due to the difference of SSP library and IMF used in Pipe3D and our work. Pipe3D adopt the \cite{Salpeter1955} IMF, while we use the \cite{Chabrier2003} IMF, which may cause a offset of 0.24 $\pm$ 0.13 dex \citep{Sanchez2016a}. For the MaNGA data, the Pipe3D pipeline use 12 SSPs from MIUSCAT library \citep{Vazdekis2012}, covering four stellar ages (0.06, 0.2, 2.0, 17.78 Gyr) and three metallicities (0.0004, 0.02, 0.0331), while we perform the spectrum fiting with 45 SSPs from \cite{Bruzual2003} model by STARLIGHT, consisting of 15 ages and 3 different metallicities (see Section \ref{subsec:spec-fitting}).  

    We use the $\ha$ emission line luminosity to determine the dust-corrected star formation rate for each spaxel, assuming a \cite{Chabrier2003} IMF and solar metallicity. The SFR is calculated from $\ha$ luminosity $L({\ha})$ using the relation from \citet{Kennicutt1998}:
    \begin{equation}
    {\rm SFR(M_{\sun} yr^{-1})} = 4.4 \ \times \ {\rm 10^{-42}} \ {\rm \times \ L({\ha})(erg \ s^{-1})}.
    \end{equation}     
    We derive the effective radius $R_e$ ($R_{50}$) and the radius enclosed 90$\%$ of light ($R_{90}$) from the NSA catalog. With the position angle (PA), $b/a$, right ascension (RA) and declination (DEC) in plate center, we also determine the deprojected galactocentric distance ($R$) for each spaxel. 

\section{Results}
\label{sec:results}

    \subsection{Metallicity Distribution In Stellar Mass -- Mass Surface Density Space}
    \label{subsec:metallicity-distribution}
\begin{figure*}
        \begin{center}
        \includegraphics[width=0.9\textwidth]{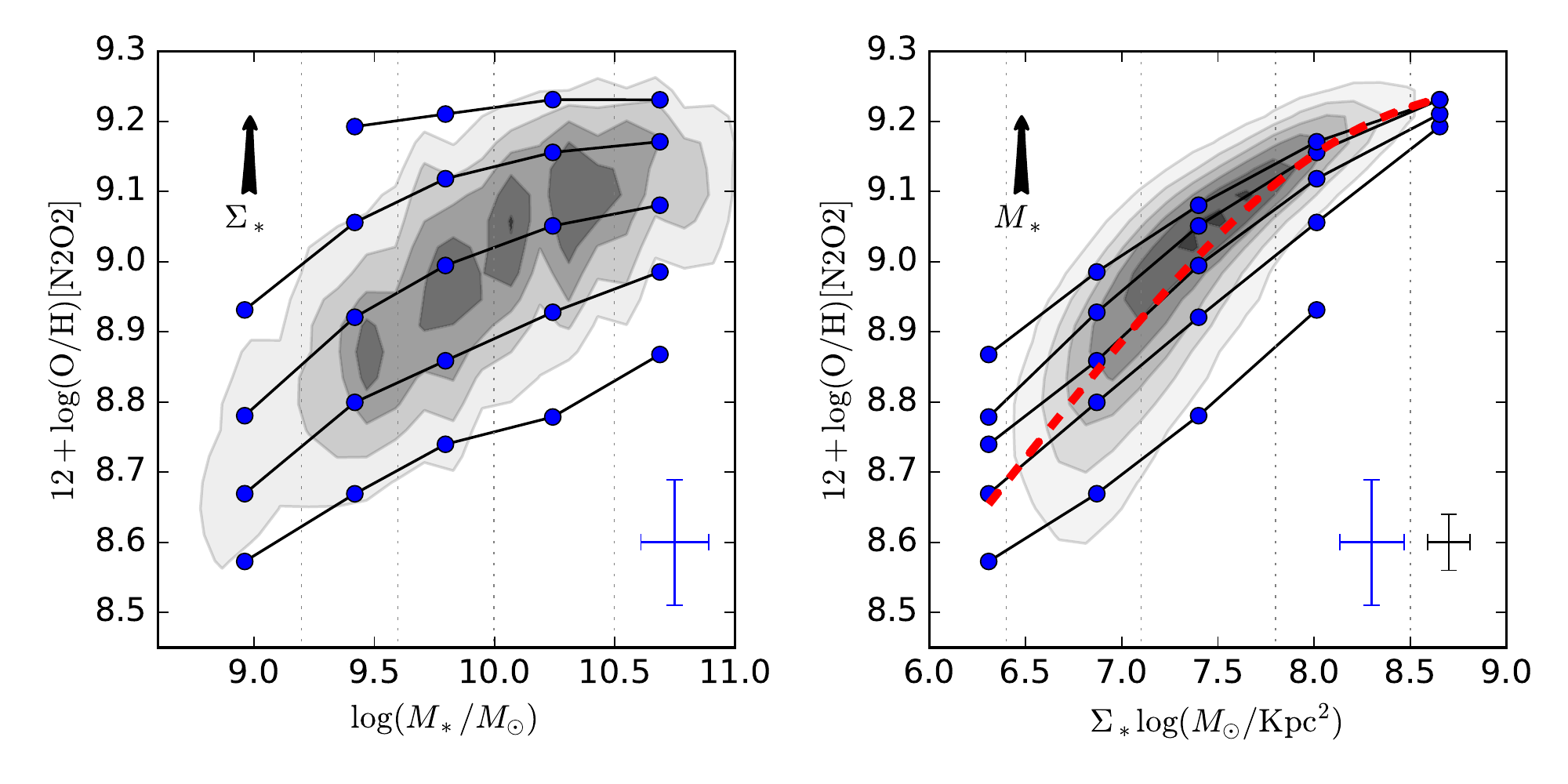}
        \end{center}
        \caption{The metallicity distribution in $M_*$ and $\Sigma_*$ space based on N2O2 index. We divide the plotting range of the logarithms of $M_*$ and $\Sigma_*$ into five bins ($M_*$: 7.8, 9.2, 9.6, 10.0, 10.5, 11.0; $\Sigma_*$: 5.7, 6.4, 7.1, 7.8, 8.5, 9.2), and thus 25 bin regions in the $M_*$ -- $\Sigma_*$ space.  The $left$ and $right$ panels show the local metallicity distribution as a function of $M_*$ and $\Sigma_*$, respectively.  Blue points connected by black lines represent the median values with different $M_*$ (or $\Sigma_*$) bins at a fixed $\Sigma_*$ (or $M_*$) bin. Bins with the number of spaxels less than 100 are not presented. The contour lines represent the distribution for $85\%$ of all the spaxels, each contour enclose the $17\%$, consecutively.  The arrow in each panel points to the direction of increasing $M_*$ or $\Sigma_*$. The black errorbar represents the typical uncertainties in the metallicity calibration with N2O2 index and mass surface density measurement, while the blue errorbars illustrate the median values of standard deviation in the 25 bin regions. The red dashed line in $\Sigma_*$ -- $Z$ panel shows the best-fitted $\Sigma_*$ -- $Z$ relation.}
        \label{fig:plot_lmass_sur_zoh}
      \end{figure*}

    This work is centered on the dependence of the local metallicity on the global stellar mass and the local stellar mass surface density. In Figure \ref{fig:plot_lmass_sur_zoh}, we plot the metallicity distribution for all the sample spaxels in star-forming regions as a function of the global stellar mass and the local stellar mass surface density, respectively. For eye-guiding purposes, we divide the plotting range of the logarithms of $M_*$ and $\Sigma_*$ into five bins ($M_*$: 7.8, 9.2, 9.6, 10.0, 10.5, 11.0; $\Sigma_*$: 5.7, 6.4, 7.1, 7.8, 8.5, 9.2), and thus 25 bin regions in the $M_*$ -- $\Sigma_*$ space. The median metallicity for the $M_*$ -- $\Sigma_*$ bins are calculated using more than 100 spaxels. 

    In Figure \ref{fig:plot_lmass_sur_zoh}, the local metallicity distribution as a function of $M_*$ and $\Sigma_*$ is shown in the $left$ and $right$ panels based on the N2O2 index. The blue points connected by black lines represent the median values with different $M_*$ (or $\Sigma_*$) bins at a fixed $\Sigma_*$ (or $M_*$) bin. The arrows in each panel point to the direction of increasing $M_*$ or $\Sigma_*$. The contour lines represent the distribution for $85\%$ of all the spaxels, each contour enclose the $17\%$, consecutively. The red dashed line in the $\Sigma_*$ -- $Z$ plot is the best-fit $\Sigma_*$ -- $Z$ relation (see details in Section \ref{subsec:relation-z-mass-surface}). 
     
    As can be seen in this figure, the metallicity increases together with $\Sigma_*$ at a certain stellar mass, and also increases with $M_*$ at a fixed surface mass density. In star-forming galaxies, the inner regions, where the surface mass density tends to be higher, are typically more metal-rich than the outskirt. This is in agreement with the previous findings that star-forming galaxies usually have negative metallicity gradients \citep[e.g.][]{Zaritsky1994,Barrera-Ballesteros2016a,Lin2017,Lian2018}. 
    
    On the other hand, for two regions with the same $\Sigma_*$ but residing in different galaxies with different $M_*$, the one in the more massive galaxy appears be more metal-rich than the other, mainly for these galaxies in the low-mass region (${\rm log}(M_*/M_\odot) \leq 10.0$), presumably indicative of the effect from global stellar mass in shaping the local metallicity.  This result has been interpreted as the evidence of metal loss by the galactic winds from galaxy potential wells in the literature, because more massive galaxies construct deeper potential wells and are less efficient in metal loss \citep[e.g.][]{Tremonti2004, Mannucci2010a, Sparre2017}. 

  \subsection{The Correlation Strength Between Stellar Mass, Stellar Mass Surface Density And Metallicity}
  \label{subsec:metallicity-correlation}
    In this section, we investigate the strength of dependence of metallicity on $M_*$ and $\Sigma_*$, and tackle the question which one is more closely related to the local metallicity in galaxies. For a given galaxy, we radially bin the spaxels so that $\Delta R/R_e=0.3$.  We use the median  values of metallicity and $\Sigma_*$ in each radial bin in the analysis below (hereafter ``radial bin sample''), though excluding the bins with less than 20 spaxels. Compared to using all the relevant spaxels, this strategy ensures an even weight for different galactic radii, and suppresses the effect of outliers. In addition, the metallicity and stellar distributions are roughly azimuthally symmetrical, thus the medians of $Z$ and $\Sigma_*$ in the radial bins appear representive. 

    Following \cite{Yang2017}, we perform a partial correlation (PCOR) analysis on our data, which is deployed to measure the correlation strength between metallicity $Z$ and $M_*$ (or $\Sigma_*$) while controlling the effects of $\Sigma_*$ (or $M_*$). There are three different statistical parameters in PCOR: one parametric statistic (Pearson) and two non-parametric statistics (Spearman and Kendall). The Pearson product-moment correlation coefficient is a measure of the linear correlation between two variables \citep{Pearson1895}, Spearman's correlation coefficient is a nonparametric measure of statistical dependence between the ranking of two variables \citep{Spearman1904}, and the Kendall correlation coefficient is used to measure the ordinal association between two measured quantities \citep{Kendall1938}. We perform the analyses with these three statistical methods on metallicity using \textsc{pcor} code \footnote{http://www.yilab.gatech.edu/pcor.html} in R language. The resultant significance ($p$ values) is listed in Table \ref{table1}. We find that all the $p$ values for $M_*$ -- $Z$ and $\Sigma_*$ -- $Z$ are large, implying that metallicity have partial dependence on both $M_*$ and $\Sigma_*$. However, we also note that all $p$ values for the $\Sigma_*$ -- $Z$ relation are slightly larger than those for the $M_*$ -- $Z$ relation, indicating that $\Sigma_*$ is more closely related to local metallicity than $M_*$. 
 
    \subsection{The Relation Between  Stellar Mass, Mass Surface Density And Metallicity }
    \label{subsec:relation-z-mass-surface}
    The global stellar mass -- metallicity relation and stellar mass -- metallicity -- SFR relation have been investigated for decades \citep[e.g.][]{Tremonti2004,Mannucci2010a,Ylgao2017}. Previous works suppress the scatter in MZR by introducing the global SFR, finding that at a certain stellar mass, the metallicity scales up remarkably when the stellar mass increases, and scales down slightly when the SFR increases. However, the scatter in the $\Sigma_*$ -- $Z$ relation remains significant (about 0.08 dex for the O3N2 estimator) in \citet{Barrera-Ballesteros2016a}. In this work, we take the global stellar mass into consideration, because the metallicity also show strong dependence on $M_*$, as can be seen in Figure \ref{fig:plot_lmass_sur_zoh}. 

    We modify the MZR relation suggested by \cite{Moustakas2011} to the following formalism: 
    \begin{equation}
    \label{eq:sigma-z-fit}
    {\rm 12+log(O/H)} = {\rm 12+log(O/H)_o} - {\mu_{\Sigma_*}} {\rm log}[1 + (\frac{\Sigma_{\rm TO}}{\Sigma_*})^{\gamma_{\Sigma_*}}],
    \end{equation} 
    where $\rm 12 + log(O/H)_o$ is the asymptotic metallicity at high stellar mass surface densities, $\Sigma_{\rm TO}$ is the turnover mass surface density, $\gamma_{\Sigma_*}$ controls the slope of the relation at low mass surface densities, and $\mu_{\Sigma_*}$ is a multiplying coefficient. We apply the Equation \ref{eq:sigma-z-fit} to fit our radial bin sample (Section \ref{subsec:metallicity-correlation}), and list the best-fit results in Table \ref{table2}. The locus of the best-fit $\Sigma_*$ -- $Z$ relation is shown with red dashed lines in Figure \ref{fig:plot_lmass_sur_zoh}. The mean value ($\mu$) and the standard deviation (scatter, $\sigma$) of the residuals when the predicted metallicity from the $\Sigma_*$ -- $Z$ relation is subtracted from the observed values derived from N2O2 index are also listed in Table \ref{table2}. $\mu_{\rm bin }$ and $\sigma_{\rm bin }$ represent the mean value and the scatter of the residuals for the radial bin sample, while $\mu_{\rm all}$ and $\sigma_{\rm all}$ are the mean value and the scatter of the residuals for all spaxels, respectively. 
     
    Since the local metallicities also strongly correlate with $M_*$ (see the upper panels in Figure \ref{fig:plot_lmass_sur_zoh}, and $p$ values for $M_*$ -- $Z$ in Table \ref{table1}), we deduce that the remarkable scatter in the $\Sigma_*$ -- $Z$ relation may be largely contributed by the variation of the stellar mass. Introducing this ingredient, we extend the $\Sigma_*$ -- $Z$ relation  (Equation \ref{eq:sigma-z-fit}) to the following format:  
    \begin{equation}
    \label{eq:fmr-fit}
    \begin{split}
    {\rm 12+log(O/H)} = {\rm 12+log(O/H)_o} - \mu_{M_*} {\rm log}[1 + (\frac{M_{\rm TO}}{M_*})^{\gamma_{M_*}}]  \\
    - \mu_{\Sigma_*} {\rm log}[1 + (\frac{\Sigma_{\rm TO}}{\Sigma_*})^{\gamma_{\Sigma_*}}], 
    \end{split}
    \end{equation}
    where $\rm 12 + log(O/H)_o$ is the asymptotic metallicity at high mass and high stellar mass surface density, $M_{\rm TO}$ is the turnover mass, $\gamma_{M_*}$ control the slope of the relation at low mass, $\mu_{M_*}$ is the coefficient. $\Sigma_{\rm TO}$, $\gamma_{\Sigma_*}$ and $\mu_{\Sigma_*}$ are the same as Equation \ref{eq:sigma-z-fit}. We apply the Equation \ref{eq:fmr-fit} to fit our radial bin sample, and list the best-fitted results in Table \ref{table2}. As shown in Table \ref{table2}, the residuals between observed metallicity and $M_*$ -- $\Sigma_*$ -- $Z$ relation for all spaxels show smaller scatter than that of the $\Sigma_*$ -- $Z$ relation as expected.

    \begin{figure*}
        \begin{center}
        \includegraphics[width=0.6\textwidth]{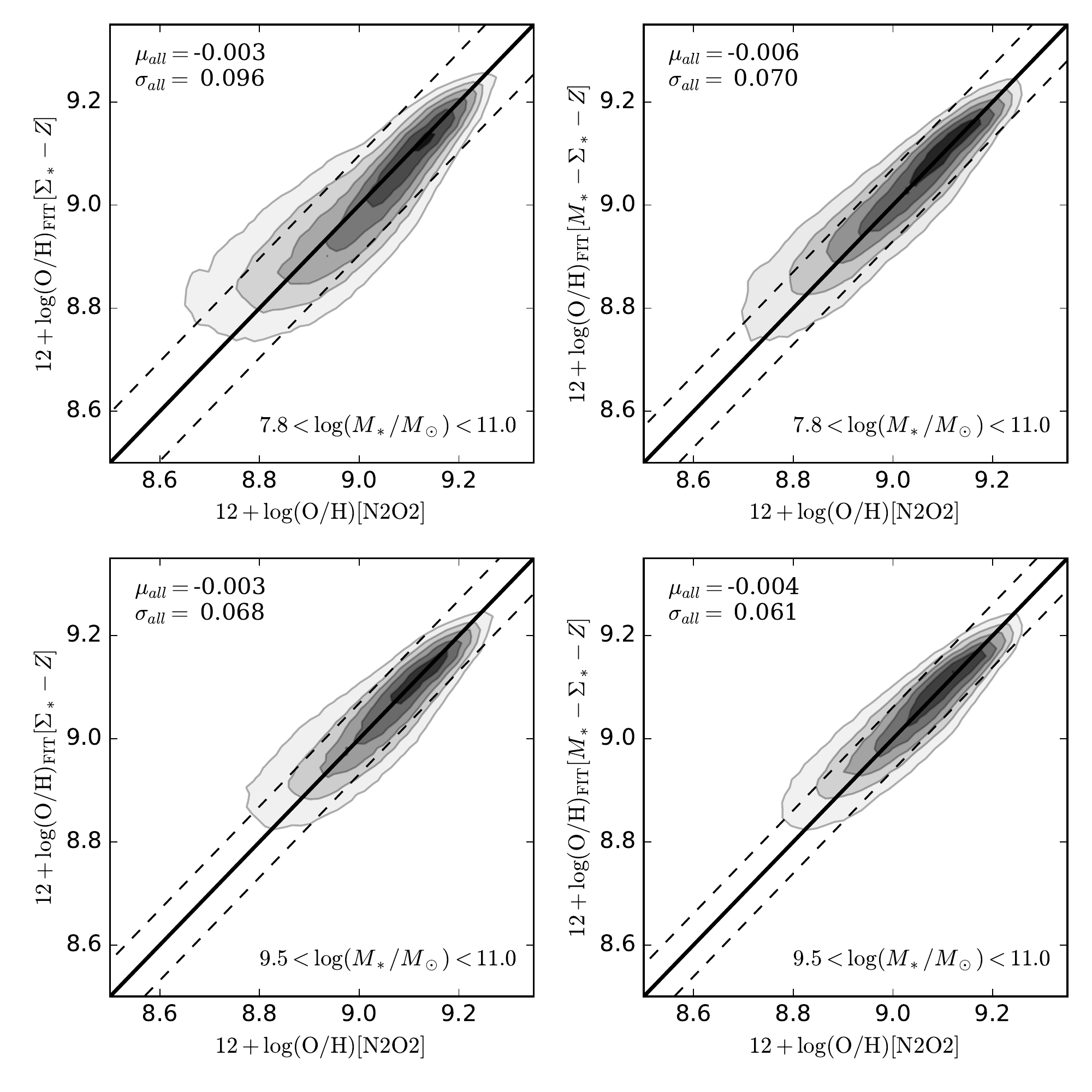}
        \end{center}
        \caption{The comparisons between the observed metallicity with N2O2 index and our best-fitted $\Sigma_*$ -- $Z$ ($left$) and $M_*$ -- $\Sigma_*$ -- $Z$ ($right$) relations for all galaxies ($top$: $7.8 < {\rm log}(M_*/M_\odot) < 11.0$) and massive galaxies ($bottom$: $9.5 < {\rm log}(M_*/M_\odot) < 11.0$ ).  The contours cover the metallicities for $85\%$ of all spaxels.  The mean value $\mu_{all}$ and standard deviation (scatter) $\sigma_{all}$ for residuals ($\Delta \rm 12+log(O/H)$) are shown in the legends. The solid lines indicate equality between the observed metallicity and best-fitted metallicity. These black dashed lines show the scatters for all spaxels, which are also shown in the legends.} 
        \label{fig:fmr_fit}
      \end{figure*}

\begin{table}[ht]
\begin{center}
\caption{$p$ values (Significances) of PCOR For $\Sigma_*$ -- $Z$ and $M_*$ -- $Z$ with N2O2 Metallicity Index. \label{table1}}
\begin{tabular}{@{}lrr@{}}
\tableline
\tableline
$p$                          & $\Sigma_*$ -- $Z$   &   $M_*$ -- $Z$          \\
\tableline
Pearson                      &   0.746           &   0.668                \\
Spearman                     &   0.798           &   0.684                \\
Kendall                      &   0.579           &   0.441                \\
\tableline
\tableline
\end{tabular}
\end{center}
\end{table}

\begin{table*}
\begin{center}
\caption{The Best-fitted Results For $\Sigma_*$ -- $Z$ and $M_*$ -- $\Sigma_*$ -- $Z$ Relations Based on N2O2 Metallicity Index.\label{table2}}
\begin{tabular}{@{}lrrlrr@{}}
\tableline
\tableline
${\rm log}(M_*/M_\odot)$   & {} & $[7.8, 11.0]$  & {} &{} & $[9.5, 11.0]$  \\ 
\cmidrule{2-3}
\cmidrule{5-6}
Parameters                   & $\Sigma_*$ -- $Z$ & $M_*$ -- $\Sigma_*$ -- $Z$  & {} &  $\Sigma_*$ -- $Z$   &   $M_*$ -- $\Sigma_*$ -- $Z$   \\
\tableline
$\rm 12+log(O/H)_o$          & 9.267 $\pm$ 0.018 &  9.320 $\pm$ 0.027 & {} & 9.267 $\pm$ 0.018 & 9.333 $\pm$ 0.032 \\
$\mu_{M_*}$                  & --                &  0.406 $\pm$ 0.107 & {} & --  & 0.824 $\pm$ 0.467 \\
${\rm log}(M_{\rm TO})$      & --                &  9.385 $\pm$ 0.140 & {} & --  & 9.016 $\pm$ 0.314 \\
$\gamma_{M_*}$               & --                &  1.062 $\pm$ 0.130 & {} & --  & 0.973 $\pm$ 0.576 \\
$ \mu_{\Sigma_*}$            & 0.350 $\pm$ 0.088 &  0.705 $\pm$ 0.302 & {} & 0.334 $\pm$ 0.095 & 0.900 $\pm$ 0.525 \\
${\rm log}(\Sigma_{\rm TO})$ & 8.050 $\pm$ 0.068 &  7.598 $\pm$ 0.349 & {} & 8.082 $\pm$ 0.088 & 7.362 $\pm$ 0.604 \\
$ \gamma_{\Sigma_*}$         & 1.010 $\pm$ 0.193 &  0.538 $\pm$ 0.126 & {} & 0.905 $\pm$ 0.189 & 0.476 $\pm$ 0.129 \\
$\mu_{\rm bin }$             & --0.0084          &  0.0001   & {} & --0.0009 & --0.0002           \\
$\sigma_{\rm bin }$          & 0.0968            &  0.0638   & {} & 0.059    & 0.052      \\
$\mu_{\rm all}$              & --0.0038          &  --0.0061 & {} & --0.003  & --0.004        \\
$\sigma_{\rm all}$           & 0.0965            &  0.0702   & {} & 0.068    & 0.061       \\
\tableline
\tableline
\end{tabular}
\begin{flushleft}
Notes: The $\rm 12+log(O/H)_o$, $\mu_{M_*}$, ${\rm log}(M_{\rm TO})$, $ \gamma_{M_*}$, $\mu_{\Sigma_*}$, ${\rm log}(\Sigma_{\rm TO})$ and $ \gamma_{\Sigma_*}$ are free parameters in $\Sigma_*$ -- $Z$ relation (Equation \ref{eq:sigma-z-fit}) and $M_*$ -- $\Sigma_*$ -- $Z$ relation (Equation \ref{eq:fmr-fit}). The $\mu_{\rm bin }$ and $\sigma_{\rm bin }$ are the mean values and scatters for distributions of residuals between our radial bin sample and $\Sigma_*$ -- $Z$  or $M_*$ -- $\Sigma_*$ -- $Z$ relation, while the $\mu_{\rm all}$ and $\sigma_{\rm all}$ are the mean values and scatters for all spaxel data. Our radial bin sample contain the median values of $\Sigma_*$, local SFRs and metallicities in the bin regions with $\Delta R/R_e=0.3$ for each galaxy, see detail in Section \ref{subsec:metallicity-correlation}. 
\end{flushleft}
\end{center}
\end{table*}

In the $top$ panels of Figure \ref{fig:fmr_fit}, we show the comparisons between the observed metallicity derived with N2O2 index and our best-fit $\Sigma_*$ -- $Z$ as well as $M_*$ -- $\Sigma_*$ -- $Z$ relations. The contours cover the metallicities for $85\%$ of all spaxels, and the solid, unity lines are overplotted for eye guidance. These black dashed lines show the scatters for all spaxels, which are 0.096 dex for $\Sigma_*$ -- $Z$ relation, while are 0.070 dex for $M_*$ -- $\Sigma_*$ -- $Z$ relation. The mean values $\mu_{all}$ and standard deviation (scatters) $\sigma_{all}$ for residuals ($\rm \Delta 12+log(O/H) = \rm 12+log(O/H) - 12+log(O/H)_{FIT}$) are shown in the legends. Note that the scatter $\sigma_{\rm all}$ in $M_*$ -- $\Sigma_*$ -- $Z$ relation is significantly smaller than the value in the $\Sigma_*$ -- $Z$ relation.  When the global stellar mass is taken into account, the scatter of the metallicity scaling relation is reduced by 27$\%$ for N2O2 index. This result means that above one quarter of the total scatter in $\Sigma_* \-- Z$ relation is due to the systematic effect with $M_*$, while the rest scatter (0.07 dex) is caused by other galactic parameters and/or intrinsic scatter (0.04 dex, N2O2) in metallicity calibration.  This tight $M_*$ -- $\Sigma_*$ -- $Z$ relation extends over three orders of magnitude in the global stellar mass, nearly four orders of magnitude in the mass surface density, a factor of 6 in metallicity with N2O2 index. 

In the $bottom$ panels of Figure \ref{fig:fmr_fit}, we  show the same relations but for massive galaxies.   We note that the scatter in  $\Sigma_*$ -- $Z$ relation is 0.068 dex, while reduced slightly to 0.061 dex in $M_*$ -- $\Sigma_*$ -- $Z$ relation.  The much smaller reduction in scatter, less than 0.01 dex, suggests that the local metallicity distribution in massive galaxies probably has no or much weak dependence on their global stellar masses. 
    
\section{Discussion}
\label{sec:discussion}

    \subsection{Comparison with Previous Studies}
    \label{subsec:comparison}
    
    In this work, we analyze the dependence of metallicity on local stellar mass surface density and global stellar mass in a sample of 1122 SFGs from the MaNGA survey.  Figure \ref{fig:plot_lmass_sur_zoh} shows that the metallicity increases largely with increasing $M_*$ (or $\Sigma_*$) at a fixed $\Sigma_*$ (or $M_*$) value. Furthermore, we perform PCOR analyses to accurately determine the strength of these correlations. From Table \ref{table1}, we note that the significance $p$ values for $M_*$ -- $Z$ and $\Sigma_*$ -- $Z$ are both prominent, indicating the importance of the roles that local $\Sigma_*$ and global stellar mass play in determining the local metallicity. However, all of the $p$ values for $\Sigma_*$ -- $Z$ are slightly larger than those for $M_*$ -- $Z$, suggesting that the local stellar mass surface density is more closely related to the local metallicity than to the galaxy stellar mass. 

    In the previous studies, some global galaxy properties, such as SFR, specific SFR, $D\rm_{n}(4000)$ and gas fraction are introduced in order to explain the large scatter in global MZR. \cite{Mannucci2010a} determined the metallicity with N2 and R23 \citep{Nagao2006} indices, and defined a new quantity $\mu_{\alpha} = {\rm log}(M_*) - \alpha {\rm log(SFR)}$ to reduce the scatter in MZR for local galaxies, resulting in a scatter of 0.05 dex in FMR. \cite{Andrews2013} found the scatter in FMR to be about 0.13 dex, where the metallicity is calculated from the electron temperature $T_e$. Similarly, for the local $\Sigma_*$ -- $Z$ relation, although \cite{Barrera-Ballesteros2016a} suggested that the $\Sigma_*$ -- $Z$ relation is largely independent of the galaxy's total stellar mass $M_*$ except at low stellar mass (${\rm log}(M_*/M_\odot) < 9.5$) and high specific SFR, they found a scatter of about 0.08 dex for the O3N2 estimator. In this work, we establish a new relation (Equation \ref{eq:sigma-z-fit}), which is a modification of the MZR in \cite{Moustakas2011}, to reinvestigate the local $\Sigma_*$ -- $Z$ relation.  We further take into account the contribution from the global $M_*$, and extend the Equation \ref{eq:sigma-z-fit} to Equation \ref{eq:fmr-fit} to reproduce the observed metallicity. As expected, the scatter in the $M_*$ -- $\Sigma_*$  -- $Z$ relation for all spaxels is significantly reduced.  However, if excluding the low-mass (${\rm log}(M_*/M_\odot) < 9.5$) galaxies in the previous analyses, we find the reductive scatter from  $\Sigma_*$ -- $Z$  to $M_*$ -- $\Sigma_*$  -- $Z$ relation is much smaller. In Figure \ref{fig:plot_residual_mass} of Appendix, we also plot the residuals between observed metallicity and the best fitted $\Sigma_*$ -- $Z$ relations as a function of stellar mass for each metallicity index. The median values of the residuals with O3N2 ($middel$ panel) are in very good agreement with the results shown in the Fig.8 of \cite{Barrera-Ballesteros2016a}. We also note that, at the low stellar mass end, the $\Sigma_*$ -- $Z$ relations with these three calibrators can not ideally reproduce the observed metallicities. As a result, for massive galaxies, our results are mostly consistent with the finding in \cite{Barrera-Ballesteros2016a}, while for low-mass galaxies, the significantly suppressed scatters,  from $\Sigma_*$ -- $Z$ to $M_*$ -- $\Sigma_*$  -- $Z$ relations, indicate that the global stellar mass has a nonnegligible effect on the local metallicity distribution of galaxies. 

    \subsection{Residuals in the $M_*$ -- $\Sigma_*$ -- $Z$ Relation}
    \label{subsec:residuals}
\begin{figure*}
        \begin{center}
        \includegraphics[width=0.9\textwidth]{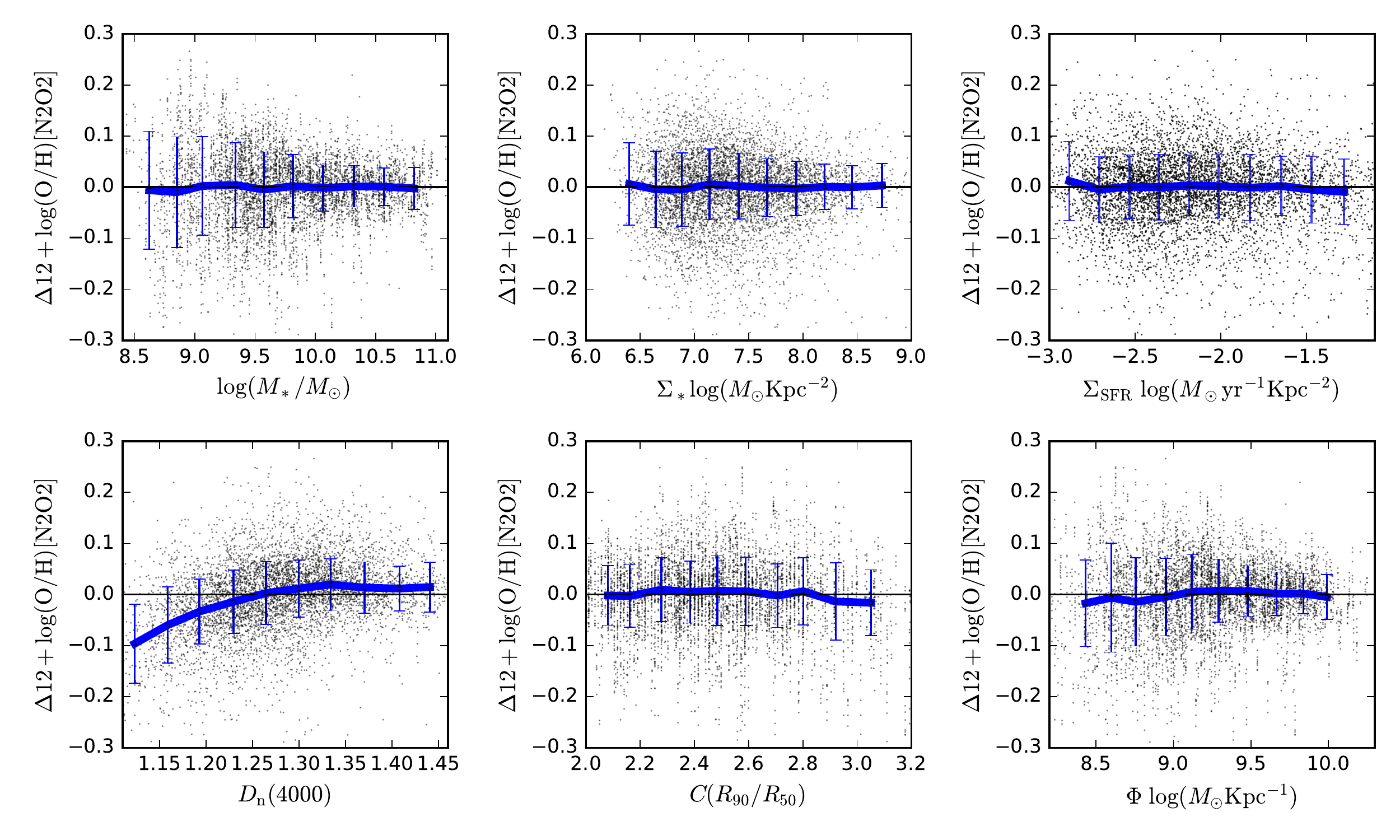}
        \end{center}
        \caption{Residuals ($\rm \Delta 12+log(O/H)$) between observed metallicities and best-fitted $M_*$ -- $\Sigma_*$ -- $Z$ relation, for radial bin sample, with respect to the total stellar mass, local stellar mass surface density, local SFR surface density, local $D_{\rm n}4000$, concentration index $C$ and average potential ($\Phi = M_* / R_e$).  The black points represent the residual values for radial bin sample, the blue line-connected points represent the median residual values for 10 bins, and corresponding errorbars represent the 16$\%$ -- 84$\%$ range in their distributions, respectively. The black solid lines represent zero-scatter in the $M_*$ -- $\Sigma_*$ -- $Z$ relation. }
        \label{fig:residuals}
      \end{figure*}

    Although the scatter in the $M_*$ -- $\Sigma_*$ -- $Z$ relation (Figure \ref{fig:fmr_fit}) is relatively small, the dependence of residuals on other galaxy properties is worth investigating, which promises to provide implications on further reducing the scatter of the metallicity scaling relations. For the radial bin sample used in Section \ref{subsec:metallicity-correlation} and Section \ref{subsec:relation-z-mass-surface}, we calculate the difference between the observed metallicity and the one derived from the best-fit curves for the $M_*$ -- $\Sigma_*$ -- $Z$ relation using N2O2 index. In Figure \ref{fig:residuals}, we show these residuals ($\rm \Delta 12+log(O/H)$) as a function of $M_*$, $\Sigma_*$, local SFR surface density ($\Sigma_{\rm SFR}$), local $D_{\rm n}4000$, the concentration index $C$ (defined as the ratio $R_{90}/R_{50}$), and the gravitational potential ($\Phi = M_* / R_e$) \citep{DEugenio2018a}. The black points represent the residuals for radial bin sample, the blue lines represent the median residual values for ten bins of each parameter, and the errorbars represent the 16$\%$ -- 84$\%$ range of the binned distributions. The black solid lines depict the zero-residual locus.

    We find that the median residuals with respect to $M_*$, $\Sigma_*$ and $\Sigma_{\rm SFR}$ approximate zero consistently, an indication that the residuals do not systematically correlate with $M_*$, $\Sigma_*$ and $\Sigma_{\rm SFR}$. The result suggests that the local metallicity is nearly independent on local SFR at a fixed $M_*$ and $\Sigma_*$, indicating that the so-called `local' FMR vanished potentially, a conclusion consistent with previous studies \citep[e.g.][]{Sanchez2013a,Barrera-Ballesteros2018}. For spiral galaxies in the local universe, \cite{Leroy2008} have demonstrated that the gas depletion time ($\Sigma_{\rm gas}/\Sigma_{\rm SFR}$) is nearly constant based on molecular and atomic gas. The growth of local stellar mass surface density is mostly the product of consumption for local gas, while the global stellar mass is also affected by the transformation in global outflow and inflow. Furthermore, for massive galaxies, the $\Sigma_* \-- Z$ relation can also successfully reproduce the global MZR, which has been studied in some works \citep[e.g.][]{Barrera-Ballesteros2016a,Sanchez2017a}, and then lead to the no or weak dependence of MZR on SFR. However, it should be noted that the hydrogen recombination lines, like $\ha$ and $\hb$, are only sensitive to the shortest timescale ($\le$ 20 Myr) of star formation \citep{Kennicutt1998}, which means that we get an instantaneous SFR estimation, instead of the previous star formation history. For massive galaxies, the majority of galaxies in this work, the local metallicity is largely independent on the local instantaneous star formation, since their long star formation history and normal star formation activities at the present day.  For low-mass metal-poor galaxies, the dependence of global MZR on SFR might appear \cite[e.g.][Gao et al. 2018, submitted]{Ly2016a}, because of their much more intense star formation activities and younger stellar populations. In the future, we will further check the existence of $M_* \-- \Sigma_* \-- Z$ relation, as well as its dependence on SFR surface density for the low-mass metal-poor SFGs based on the IFS data.  

    However, the residuals seemly show a weak correlation with $D_{\rm n}4000$. With increasing $D_{\rm n}4000$, the absolute values of residuals decrease when $D_{\rm n}4000<$ 1.35. The residuals are negative when $D_{\rm n}4000 < 1.25$, and are positive when $1.25 < D_{\rm n} 4000 < 1.50$. These results indicate that our $M_*$ -- $\Sigma_*$ -- $Z$ relation overestimate the metallicity for very young ($\lesssim$ 0.25 Gyr) galactic regions \citep{Kauffmann2003}, but underestimate the metallicity for galactic regions with an older stellar age. \cite{Lian2015} found that the metallicity is positive correlated to $D\rm_{n}(4000)$, which indicates that galaxies with older stellar ages have higher metallicities. Furthermore, we also find that the residuals for concentration index $C$ and gravitational potential $\Phi$ are nearly zero, meaning that the local metallicity may be independent on the galaxy concentration and average gravitational potential. 
    Recently, \cite{wang2017} have demonstrated that the local metallicity also depends on the assembly modes of galaxies. In the sense that higher $\Sigma_*$ (denser) regions, located in the inner regions of galaxies with an ``outside-in'' assembly mode, have higher metallicity and lower $D_{\rm n}4000$ than those galaxies with an ``inside-out'' mode. This may lead to additional scatter in $M_*$ -- $\Sigma_*$  -- $Z$ relation. In the future, we expect to further reduce the scatter in $M_*$ -- $\Sigma_*$  -- $Z$ relation by considering the different assembly modes of galaxies. 

    \subsection{The Impact of Metallicity Calibrators}
    \label{subsec:metal_calib}
    Apart from the N2O2 index used in the main text, we also estimate the metallicity with O3N2 and N2 diagnostics (see Appendix) to check whether our primary results are independent on the different metallicity estimators. The O3N2 and N2 calibrators are anchored by the ``direct method'' with electron temperature proposed by \cite{Marino2013}, while the N2O2 calibrator is constructed by comparing the line ratios with photo-ionization models \citep{Kewley2002}. However, it is well known that these two procedures will lead to a relatively nonnegligible difference in metallicity estimation \citep{Morisset2016}.  \cite{Pettini2004} and \cite{Lopez-Sanchez2011} have clearly shown that, when using the O3N2 and N2 diagnostics, the regions with higher ionization degree tend to have lower oxygen abundances. In particular, the N2O2 diagnostic is subject to variations in nitrogen-to-oxygen abundance ratio (N/O) ratios and temperature, and is sensitive to metallicity only for $Z>8.3$ \citep{Dopita2000,Dopita2013,Perez-Montero2016}.  

    In Figure \ref{fig:plot_lmass_sur_zoh} and \ref{fig:plot_lmass_sur_zoh_o3n2_n2}, we note that the metallicity with N2O2 index covers a wide range about 0.6 dex, while a narrow range less than 0.3 dex with N2 index, since the saturation that N2 suffered in the high-metallicity region. In addition, as shown in Figure \ref{fig:plot_residual_mass}, the absolute values of residual values increase with decreasing stellar mass, regardless the calibrator, while the absolute values of the residuals using O3N2 and N2 are systematically smaller than the N2O2 calibrator. Despite this, we find the trend, that the local metallicity increases with $M_*$ (or $\Sigma_*$) at a fixed $\Sigma_*$ (or $M_*$), is similar to the result with N2O2 index. In Figure \ref{fig:plot_fmr_fit_o3n2} and \ref{fig:plot_fmr_fit_n2}, the scatters are also reduced by about 19$\%$ and 25$\%$ from $\Sigma_*$  -- $Z$ to $M_*$ -- $\Sigma_*$  -- $Z$ relations with O3N2 and N2 indices, respectively. Nevertheless, if excluding these low-mass galaxies, the reduced scatters are much smaller, which is consistent with the result with N2O2 index. 

   	\subsection{Implications of the $M_*$ -- $\Sigma_*$ -- $Z$ Relation}
   	\label{subsec:interpretation}
   	In Section \ref{sec:results}, we find that a tight correlation exists between the global stellar mass, the local stellar mass surface density and the local metallicity, especially for low-mass star-forming galaxies, in which the metallicity increases with $M_*$ and $\Sigma_*$ in a systematic way. The local metallicities are mainly determined by the metals produced in the past, the metal loss by galactic outflows \citep{Barrera-Ballesteros2018} or AGN feedback \citep{Wylezalek2017}, and the metal dilution by cold gas inflows. The local stellar mass surface density is a result of the local galactic stellar assembly history \citep{Yozin2016,Goddard2017,Goddard2017a,Jones2017}, the strong correlation for $\Sigma_*$ -- $Z$ can be naturally explained by noticing that the higher $\Sigma_*$ (denser) regions usually locate in the inner region of galaxies \citep{Johnston2017}, which have longer star formation history than the less dense regions with lower $\Sigma_*$ \citep{Ibarra-Medel2016}. 	
   	
   	Furthermore, the global stellar mass reflects the assembly of the stellar mass and the depth of the potential wells, which dominates the infall of metal-poor gas and the outflows of metal-rich gas \citep{Peeples2011,Pan2015,Cheung2016,Lian2018}. As proposed by \cite{Tremonti2004} and \cite{Lian2018}, the deep gravitational potential wells of massive galaxies retain the processed metals, leading to a higher local metallicity than less massive galaxies even at a similar $\Sigma_*$. In general, less massive galaxies are expected to be more efficient in diluting the metal-rich gas, caused by metal-poor gas inflows, and removing the metal-rich gas by stronger galactic outflows \citep{Chisholm2018}, due to their shallower gravitational potential wells and longer timescale for the inflow.  Compared with these high-mass galaxies, low-mass galaxies are less efficient in converting gas into stars, which is the so-called `downsizing' scenario \citep{Thomas2010}, and thus have the relatively higher gas fraction ($\Sigma_{\rm gas}/\Sigma_*$). \cite{Barrera-Ballesteros2018} also argued that the gas fraction decreases tightly with the increasing local stellar mass surface density.  For galaxies with higher stellar mass and higher local stellar mass surface density, the star formation activity becomes relatively weaker because of the lower gas fraction. Recently, \cite{Chisholm2018} has suggested that it is easier to significantly remove the metals by stronger galactic outflows in low-mass galaxies than in massive galaxies. Nevertheless, we should notice that \cite{Chisholm2018} just derived the outflow properties for seven galaxies, which cover a wide range of stellar mass but lack the range of $7.5 \le {\rm log}(M_*/M_\odot) \le 9.0$.  In brief, the dependence of local stellar mass surface density -- local metallicity relation on global stellar mass is pronounced for low-mass galaxies, since their shallower gravitational potential wells, while is more or less weaker for massive galaxies.   

\section{Summary} 
\label{sec:summary}
    In this work, we have used the IFS data from the MaNGA survey to investigate the global stellar mass -- local stellar mass surface density -- metallicity relation, and the strengths of their correlation with metallicity in star-forming galaxies. In total, we select 1122 star-forming galaxies with about 750,000 spaxels in star-forming regions as our sample. We have determined the local metallicities with the N2O2 diagnostic. The local stellar mass surface densities are derived from the best-fit results given by \textsc{STARLIGHT}, and the SFR surface density are estimated from $\ha$ luminosities. The main results and conclusions of this paper are summarized as follows.

    \begin{itemize}
      \item  In agreement with previous studies, the metallicity increases steeply with increasing surface stellar mass density at a fixed stellar mass. Similarly, at a fixed surface stellar mass density, the metallicity increases strongly with increasing stellar mass at low mass end, while this trend becomes less obvious at high mass end (Figure \ref{fig:plot_lmass_sur_zoh}).  Performing partial correlation analyses on $M_*$,  $\Sigma_*$ and $Z$ (Table \ref{table1}), we find that the $p$ values for the correlations of $M_*$ -- $Z$  and $\Sigma_*$ -- $Z$ to be large, indicating that metallicity is partially dependent on both  global stellar mass and local  mass  surface density.  However, all $p$ values for the  $\Sigma_*$ -- $Z$ relation are slightly larger than those for the $M_*$ -- $Z$  relation, indicating that the local metallicity $Z$ correlates with $\Sigma_*$ more strongly than $M_*$. 

      \item  We establish the $\Sigma_*$ -- $Z$ relation (Equation \ref{eq:sigma-z-fit}) following the relation in \cite{Moustakas2011}, and extend the $\Sigma_*$ -- $Z$ relation to a new $M_*$ -- $\Sigma_*$ -- $Z$ relation (Equation \ref{eq:fmr-fit}, Figure \ref{fig:fmr_fit}). Compared to $\Sigma_*$ -- $Z$ relation,  the scatter in the $M_*$ -- $\Sigma_*$ -- $Z$ relation is reduced by about 30$\%$ for galaxies with $7.8 < {\rm log}(M_*/M_\odot) < 11.0$, while the reduced scatter is much smaller when excluding the low-mass (${\rm log}(M_*/M_\odot) < 9.5$) galaxies, suggesting that the $M_*$ -- $\Sigma_*$ -- $Z$ relation is a more universal and fundamental relation than $M_*$ -- $Z$ and $\Sigma_*$ -- $Z$ relations for those low-mass galaxies. 

      \item  We find the residuals in the best-fitted $M_*$ -- $\Sigma_*$ -- $Z$ relation to be possibly correlated with the $D_{\rm n}4000$ (Figure \ref{fig:residuals}). When including the contribution of the SFR, we find that the local metallicity is largely independent on the local SFR surface density at a fixed $M_*$ and $\Sigma_*$, indicating the lack of so-called `local' FMR, consistent with previous studies.
    \end{itemize} 

    We emphasize that our basic results do not depend on the different metallicity estimators (e.g. O3N2, N2, see Appendix).  The local metallicity can be determined well with the global stellar mass and local stellar mass surface density, suggesting that the local metallicity is primarily determined by both the local galactic stellar mass assembly history and the global stellar mass assembly history. We interpret our result as the combination of the produced metals in the local star formation history and the metal loss due to the galactic  winds from the galactic potential wells.  Furthermore, the remaining scatter in the $M_*$ -- $\Sigma_*$ -- $Z$ relation may be contributed by the local $D_{\rm n}4000$.  

\acknowledgments

We thank the referee for her/his constructive comments. We thank Jianhui Lian for useful discussion on the manuscript.
This work is supported by the National Key R\&D Program of China (2015CB857004, 2016YFA0400702, 2017YFA0402600), and the National Natural Science Foundation of China (NSFC, Nos. 11320101002, 11421303, and 11433005). 
Enci Wang acknowledges the support from the Youth Innovation Fund by University of Science and Technology of China (No. WK2030220019) and the China Postdoctoral Science Foundation funded project (No. BH2030000040). Guilin Liu acknowledges the support from the National Thousand Young Talents Program of China.

Funding for the Sloan Digital Sky Survey IV has been provided by the Alfred P. Sloan Foundation, the U.S. Department of Energy Office of Science, and the Participating Institutions. SDSS acknowledges support and resources from the Center for High-Performance Computing at the University of Utah. The SDSS web site is www.sdss.org.

SDSS is managed by the Astrophysical Research Consortium for the Participating Institutions of the SDSS Collaboration including the Brazilian Participation Group, the Carnegie Institution for Science, Carnegie Mellon University, the Chilean Participation Group, the French Participation Group, Harvard-Smithsonian Center for Astrophysics, Instituto de Astrofísica de Canarias, The Johns Hopkins University, Kavli Institute for the Physics and Mathematics of the Universe (IPMU) / University of Tokyo, Lawrence Berkeley National Laboratory, Leibniz Institut für Astrophysik Potsdam (AIP), Max-Planck-Institut für Astronomie (MPIA Heidelberg), Max-Planck-Institut für Astrophysik (MPA Garching), Max-Planck-Institut für Extraterrestrische Physik (MPE), National Astronomical Observatories of China, New Mexico State University, New York University, University of Notre Dame, Observatório Nacional / MCTI, The Ohio State University, Pennsylvania State University, Shanghai Astronomical Observatory, United Kingdom Participation Group, Universidad Nacional Autónoma de México, University of Arizona, University of Colorado Boulder, University of Oxford, University of Portsmouth, University of Utah, University of Virginia, University of Washington, University of Wisconsin, Vanderbilt University, and Yale University. 

\bibliography{ms}

\newpage
\begin{appendix}

\section{The metallicity calculation with O3N2 and N2 diagnostics}
\label{appendix}

    The diagnositc O3N2 index \citep{Alloin1979} is defined as 
    \begin{equation}
    \rm O3N2 \equiv log(\frac{\oiii\lambda5007}{\hb} \times \frac{\ha}{\nii\lambda6583}).
    \end{equation}    
    The empirical calibration of O3N2 diagnostic for the specific purpose of calculating metallicity, improved by \cite{Marino2013}, is 
    \begin{equation}
    \rm 12 + log(O/H) = 8.505 - 0.221 \times O3N2  
    \end{equation}
    with O3N2 ranging from $-1.1$ to 1.7. The typical error for the metallicity calibration with the O3N2 diagnostic is 0.08 dex.
    
    The diagnostic N2 index \citep{Storchi-Bergmann1994,Raimann2000} is defined as 
    \begin{equation}
    \rm N2 \equiv log(\nii\lambda6583 / \ha),
    \end{equation}
    and the metallicity calibration relation, derived by \cite{Marino2013}, is given by
    \begin{equation}
    \rm 12 + log(O/H) = 8.667 + 0.455 \times N2, 
    \end{equation} 
    with an average uncertainty of 0.09 dex, where the N2 index is between $-1.6$ and $-0.2$.

\section{The pcor analysis, $\Sigma_*$ -- $Z$ and $M_*$ -- $\Sigma_*$ -- $Z$ relations with O3N2 and N2 indices}

\begin{table}[ht]
\begin{center}
\caption{$p$ values (Significances) of PCOR Based on O3N2 and N2 Metallicity Indices.\label{table3}}
\begin{tabular}{@{}lrrrrr@{}}
\tableline
\tableline
$p$                          & O3N2              & {}    & {}          & N2        & {} \\ 
\cmidrule{2-3}
\cmidrule{5-6}         
{}                           & $\Sigma_*$ -- $Z$   & $M_*$ -- $Z$ & {}  & $\Sigma_*$ -- $Z$ &  $M_*$ -- $Z$  \\
\tableline                               
Pearson                      &   0.699           & 0.603    & {}    & 0.665           & 0.627            \\
Spearman                     &   0.746           & 0.613    & {}    & 0.704           & 0.680           \\
Kendall                      &   0.545           & 0.405    & {}    & 0.485           & 0.462             \\
\tableline
\tableline
\end{tabular}
\end{center}
\end{table}

\begin{table}[ht]\tiny
\begin{center}
\caption{The Best-fitted Results For $\Sigma_*$ -- $Z$ and $M_*$ -- $\Sigma_*$ -- $Z$ Relations Based on O3N2 and N2 Metallicity Indices.\label{table4}}
\begin{tabular}{@{}lrrrrrlrrrrr@{}}
\tableline
\tableline
Parameters  & {} & O3N2  & {}  & {} & {}  & {}  & {} & N2 & {} & {} \\ 
\cmidrule{2-6}
\cmidrule{8-12}
${\rm log}(M_*/M_\odot)$   & {} & $[7.8, 11.0]$  & {} &{} & $[9.5, 11.0]$ &
{}   & {} & $[7.8, 11.0]$  & {} &{} & $[9.5, 11.0]$  \\
\cmidrule{2-3}
\cmidrule{5-6}
\cmidrule{8-9}
\cmidrule{11-12}
{}                   & $\Sigma_*$ -- $Z$   & $M_*$ -- $\Sigma_*$ -- $Z$ & {} & $\Sigma_*$ -- $Z$   & $M_*$ -- $\Sigma_*$ -- $Z$ & {} & $\Sigma_*$ -- $Z$   & $M_*$ -- $\Sigma_*$ -- $Z$ & {} & $\Sigma_*$ -- $Z$   & $M_*$ -- $\Sigma_*$ -- $Z$   \\
\tableline
$\rm 12+log(O/H)_o$          & 8.575 $\pm$ 0.009 & 8.565 $\pm$ 0.009 & {} & 8.575$\pm$0.009 & 8.558$\pm$0.007 & {}
                             & 8.595 $\pm$ 0.009 & 8.571 $\pm$ 0.003 & {} & 8.590$\pm$0.008 & 8.568$\pm$0.005 \\
$\mu_{M_*}$                  & --                & 0.148 $\pm$ 0.044 & {} & -- & 0.397$\pm$0.290 & {}
                             & --                & 0.249 $\pm$ 0.050 & {} & -- & 0.222$\pm$0.668 \\
${\rm log}(M_{\rm TO})$      & --                & 9.523 $\pm$ 0.133 & {} & -- & 9.083$\pm$0.306 & {}
                             & --                & 9.310 $\pm$ 0.122 & {} & -- & 9.362$\pm$1.876 \\
$\gamma_{M_*}$               & --                & 1.190 $\pm$ 0.189 & {} & -- & 1.261$\pm$0.772 & {}
                             & --                & 0.898 $\pm$ 0.085 & {} & -- & 0.913$\pm$0.553 \\
$ \mu_{\Sigma_*}$            & 0.265 $\pm$ 0.068 & 0.230 $\pm$ 0.077 & {} & 0.250$\pm$0.070 & 0.302$\pm$0.100 & {}
                             & 0.345 $\pm$ 0.146 & 0.102 $\pm$ 0.018 & {} & 0.300$\pm$0.164 & 0.139$\pm$0.037 \\
${\rm log}(\Sigma_{\rm TO})$ & 7.755 $\pm$ 0.097 & 7.669 $\pm$ 0.165 & {} & 7.738$\pm$0.117 & 7.392$\pm$0.204 & {}
                             & 7.315 $\pm$ 0.275 & 7.361 $\pm$ 0.052 & {} & 7.214$\pm$0.410 & 7.170$\pm$0.104 \\
$ \gamma_{\Sigma_*}$         & 0.925 $\pm$ 0.163 & 0.801 $\pm$ 0.167 & {} & 0.905$\pm$0.163 & 0.799$\pm$0.136 & {}
                             & 0.75 $\pm$ 0.161 & 1.572 $\pm$ 0.186  & {} & 0.702$\pm$0.167 & 1.439$\pm$0.191 \\
$\mu_{\rm bin }$             & -0.005          & 0.0004              & {} & --0.003 & 0.003 & {}
                             & 0.0026            & 0.0022            & {} & --0.002 & --0.0001 \\
$\sigma_{\rm bin }$          & 0.0528            & 0.0390            & {} & 0.0358 & 0.0332 & {}
                             & 0.0467            & 0.0320            & {} & 0.0309 & 0.0252 \\
$\mu_{\rm all}$              & 0.002          & 0.0018               & {} & 0.0006 & 0.0027 & {}
                             & 0.0011            & --0.0012          & {} & --0.003 & -0.002 \\
$\sigma_{\rm all}$           & 0.0571            & 0.0465            & {} & 0.0452 & 0.0424 & {}
                             & 0.0515            & 0.0385            & {} & 0.0385 & 0.0332  \\
\tableline
\tableline
\end{tabular}
\end{center}
\end{table}

\begin{figure*}[ht]
        \begin{center}
        \includegraphics[width=0.6\textwidth]{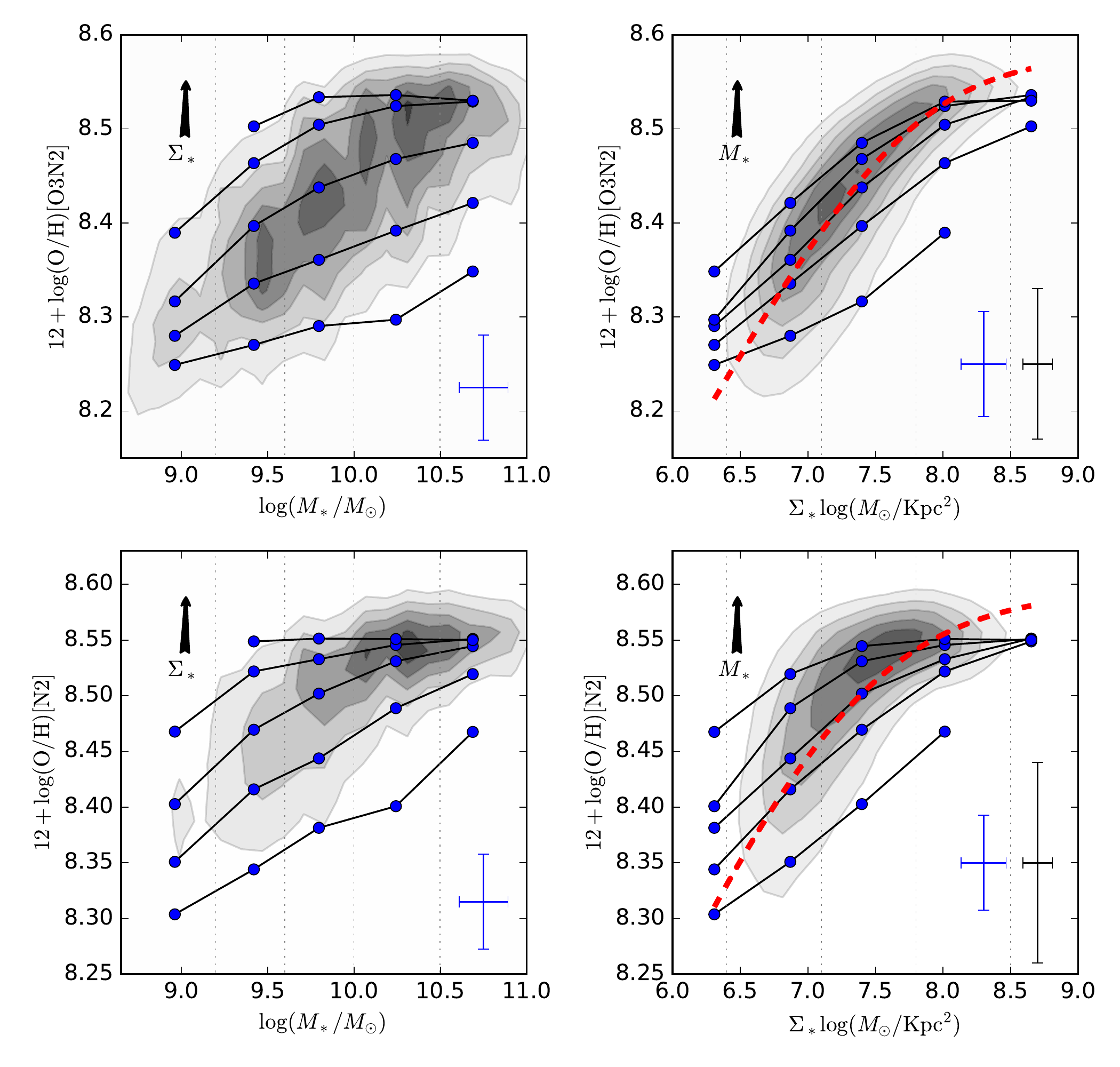}
        \end{center}
        \caption{The metallicity distribution in $M_*$ and $\Sigma_*$ space based on O3N2 and N2 indices.   The meaning of symbols is same as Figure \ref{fig:plot_lmass_sur_zoh}. }
        \label{fig:plot_lmass_sur_zoh_o3n2_n2}
\end{figure*}

\begin{figure*}[ht]
        \begin{center}
        \includegraphics[width=0.9\textwidth]{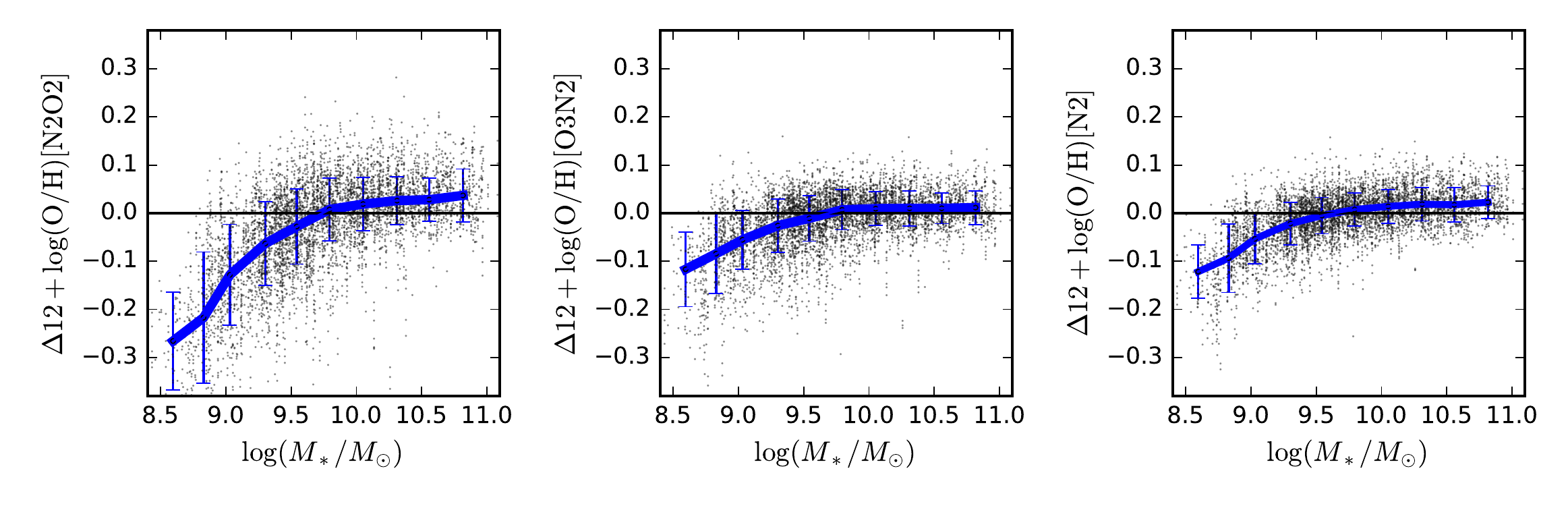}
        \end{center}
        \caption{Residuals ($\rm \Delta 12+log(O/H)$) between observed metallicities and best-fitted $\Sigma_*$ -- $Z$ relation, for three metallicity calibrators, with respect to the total stellar mass. The black points represent the residual values for radial bin sample, the blue line-connected points represent the median residual values for 10 bins, and corresponding errorbars represent the 16$\%$ -- 84$\%$ range in their distributions, respectively. The black solid lines represent zero-scatter in the $\Sigma_*$ -- $Z$ relation. }
        \label{fig:plot_residual_mass}
\end{figure*}

\begin{figure*}[ht]
        \begin{center}
        \includegraphics[width=0.5\textwidth]{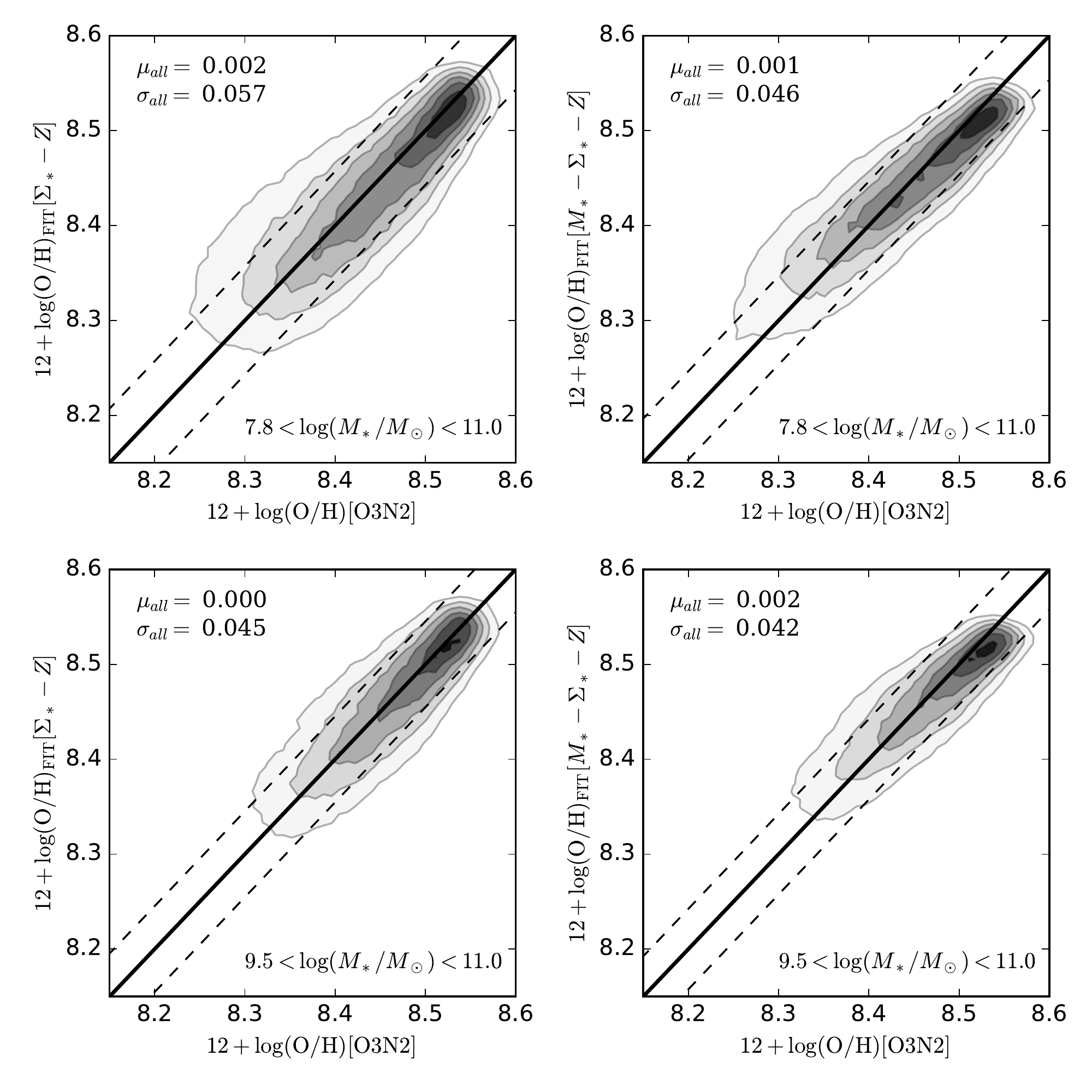}
        \end{center}
        \caption{The comparisons between the observed metallicity with O3N2 index and our best-fitted $\Sigma_*$ -- $Z$ ($left$) and $M_*$ -- $\Sigma_*$ -- $Z$ ($right$) relations for all galaxies ($top$: $7.8 < {\rm log}(M_*/M_\odot) < 11.0$) and massive galaxies ($bottom$: $9.5 < {\rm log}(M_*/M_\odot) < 11.0$ ). The meaning of symbols is same as Figure \ref{fig:fmr_fit}. }
        \label{fig:plot_fmr_fit_o3n2}
\end{figure*}

\begin{figure*}[ht]
        \begin{center}
        \includegraphics[width=0.5\textwidth]{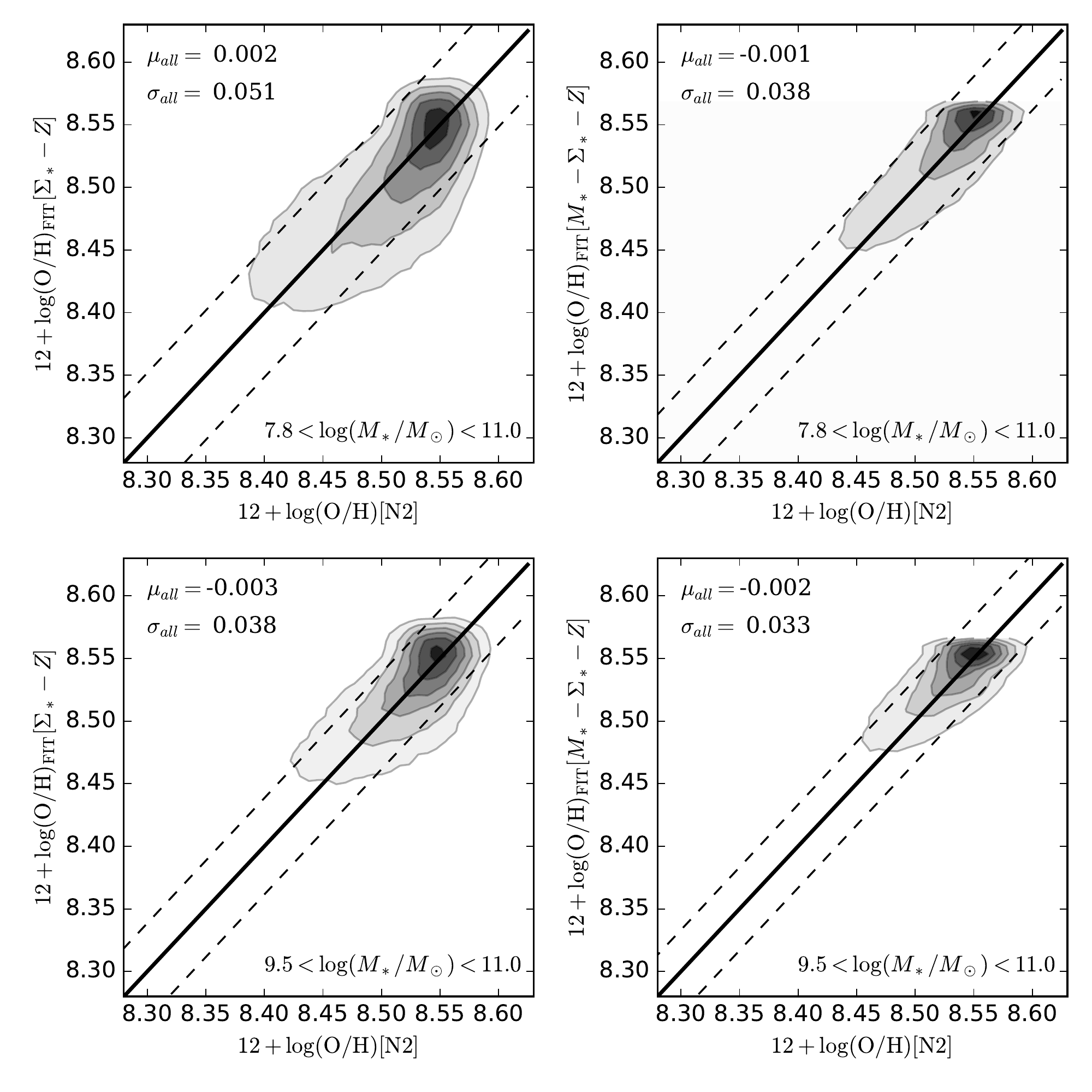}
        \end{center}
        \caption{The comparisons between the observed metallicity with  N2 index and our best-fitted $\Sigma_*$ -- $Z$ ($left$) and $M_*$ -- $\Sigma_*$ -- $Z$ ($right$) relations for all galaxies ($top$: $7.8 < {\rm log}(M_*/M_\odot) < 11.0$) and massive galaxies ($bottom$: $9.5 < {\rm log}(M_*/M_\odot) < 11.0$ ). The meaning of symbols is same as Figure \ref{fig:fmr_fit}. }
        \label{fig:plot_fmr_fit_n2}
\end{figure*}

\end{appendix}

\end{document}